\documentclass[twocolumn]{emulateapj}

\usepackage{epsfig}
\usepackage{graphicx}
\usepackage{amsmath, amsthm, amssymb}
\usepackage{rotating}
\usepackage{multirow}

\shortauthors{Lim et al.}

\begin{document}

\title{The Distribution of Mass Surface Densities in a High-Mass Protocluster}

\author{Wanggi Lim\altaffilmark{1}, Jonathan C. Tan\altaffilmark{1,2}, Jouni Kainulainen\altaffilmark{3}, Bo Ma\altaffilmark{1} and Michael J. Butler\altaffilmark{3}}
\affil{\footnotesize \altaffilmark{1}Department. of Astronomy, University of Florida, Gainesville, FL 32611, USA}
\affil{\footnotesize \altaffilmark{2}Department of Physics, University of Florida, Gainesville, FL 32611, USA}
\affil{\footnotesize \altaffilmark{3}Max-Planck-Institute for Astronomy, K{\"o}nigstuhl 17, 69117 Heidelberg, Germany}

\begin{abstract}
We study the probability distribution function (PDF) of mass surface
densities, $\Sigma$, of infrared dark cloud (IRDC) G028.37+00.07 and
its surrounding giant molecular cloud. This PDF constrains the
physical processes, such as turbulence, magnetic fields and
self-gravity, that are expected to be controlling cloud structure and
star formation activity. The chosen IRDC is of particular interest
since it has almost 100,000 solar masses within a radius of 8 parsecs,
making it one of the most massive, dense molecular structures known
and is thus a potential site for the formation of a ``super star
cluster.'' We study $\Sigma$ in two ways. First, we use a combination
of NIR and MIR extinction maps that are able to probe the bulk of the
cloud structure up to
$\Sigma\sim1\:{\rm{g\:cm}^{-2}}\:$($A_V\simeq200$~mag). Second, we
study the FIR and sub-mm dust continuum emission from the cloud
utilizing Herschel PACS and SPIRE images and paying careful attention
to the effects of foreground and background contamination. We find
that the PDFs from both methods, applied over a
$\sim20^\prime$(30~pc)-sized region that contains
$\simeq1.5\times10^5\:M_\odot$ and encloses a minimum closed contour
with $\Sigma\simeq0.013\:{\rm{g\:cm}^{-2}}\:$($A_V\simeq3$~mag), shows
a log-normal shape with the peak measured at
$\Sigma\simeq0.021\:{\rm{g\:cm}^{-2}}\:$($A_V\simeq4.7$~mag).  There
is tentative evidence for the presence of a high-$\Sigma$ power law
tail that contains from $\sim3\%$ to 8\% of the mass of the cloud
material.  We discuss the implications of these results for the
physical processes occurring in this cloud.
\end{abstract}


\section{Introduction}

The probability distribution function (PDF) of mass surface density,
$\Sigma$, is one of the simplest metrics of interstellar cloud
structure. This $\Sigma$-PDF is, in principle, much easier to observe
than other distributions, such as volume density, thus making it a
convenient metric with which to compare observed and simulated clouds.
The $\Sigma$-PDF shape should be sensitive to physical processes
occurring in the clouds. For example, simulations of driven supersonic
hydrodynamic (and if including magnetic fields, super-Alfv\'enic)
turbulence of non-self-gravitating gas in periodic boxes yield
lognormal $\Sigma$-PDFs (e.g., Federrath 2013; Padoan et
al. 2014), i.e., the area-weighted PDF, $p_A$, can be well-fit by a
lognormal:
\begin{equation}
p_A({\rm{ln}}\Sigma^\prime)=\frac{1}{(2\pi)^{1/2}\sigma_{\rm{ln}\Sigma^\prime}}{\rm{exp}}\left[-\frac{({\rm{ln}}\Sigma^{\prime}-\overline{{\rm{ln}}\Sigma})^2}{2\sigma_{\rm{ln}\Sigma^\prime}^2}\right],
\label{eq:lognormal}
\end{equation}
where $\Sigma^\prime\equiv\Sigma/\overline{\Sigma}_{\rm{PDF}}$ is
mean-normalized $\Sigma$. The lognormal width,
$\sigma_{\rm{ln}\Sigma}$, grows as turbulent Mach number increases. In
simulations with self-gravity, $\Sigma$-PDFs are seen to develop
high-end power law tails, perhaps tracing regions undergoing free-fall
collapse (Kritsuk et al. 2011; Collins et al. 2011; Cho \& Kim 2011;
Federrath \& Klessen 2013). However, simulations of self-gravitating,
strongly-magnetized (trans-Alfv\'enic), turbulent clouds with
non-periodic boundary conditions are also needed for comparison with
observed $\Sigma$-PDFs. Such clouds are expected to have smaller star
formation efficiencies per mean free-fall time, $\epsilon_{\rm{ff}}$,
likely implying they would have smaller mass fractions in any
high-$\Sigma$ power law tail. Accurate quantification of the
$\Sigma$-PDF in real star-forming clouds is needed to constrain
theoretical models.

Observationally, Kainulainen et al. (2009 [K09]) performed NIR extinction
mapping of $\sim$20 nearby clouds, ranging from ``quiescent,''
non-star-forming clouds to more active clouds. They found quiescent
cloud $\Sigma$-PDFs are well-described by lognormals, while
star-forming clouds have high-$\Sigma$ power law tails. Note, in
practice a lognormal is fit to the observed $p_A({\rm{ln}}\Sigma)$,
which is then used to derive $\overline{\rm{ln}\Sigma}$ (over the
considered range of $\Sigma$), which then defines the mean
$\overline{\Sigma}_{\rm{PDF}}\equiv{e}^{\overline{\rm{ln}\Sigma}+\sigma^2_{\rm{ln}\Sigma}/2}$,
where $\sigma_{\rm{ln}\Sigma}$ is standard deviation of
$\rm{ln}\Sigma$ (Butler et al. 2014, hereafter BTK14).

However, the ability of this observational method to accurately
measure the position of the $\Sigma$-PDF peak, typically at
$\Sigma\sim0.01\:{\rm{g\:cm}^{-2}}$ (i.e., $A_V\sim2\:{\rm{mag}}$: we
adopt conversion
$\Sigma/(1\:{\rm{g\:cm}^{-2}})\equiv{A}_v/(224.8\:{\rm{mag}})$
(Kainulainen \& Tan 2013 [KT13]) in the K09 clouds, has been
questioned by Schneider et al. (2015a) and Lombardi et al. (2015) due
to difficulties of disentangling foreground and background
contributions. Lombardi et al. argued that low-$\Sigma$ PDF
uncertainties are so large that observed PDFs are all consistent
with power laws.

\begin{figure*}
\centering
\begin{tabular}[b]{c@{\hspace{-0.08in}}c}
\includegraphics[width=5.5in]{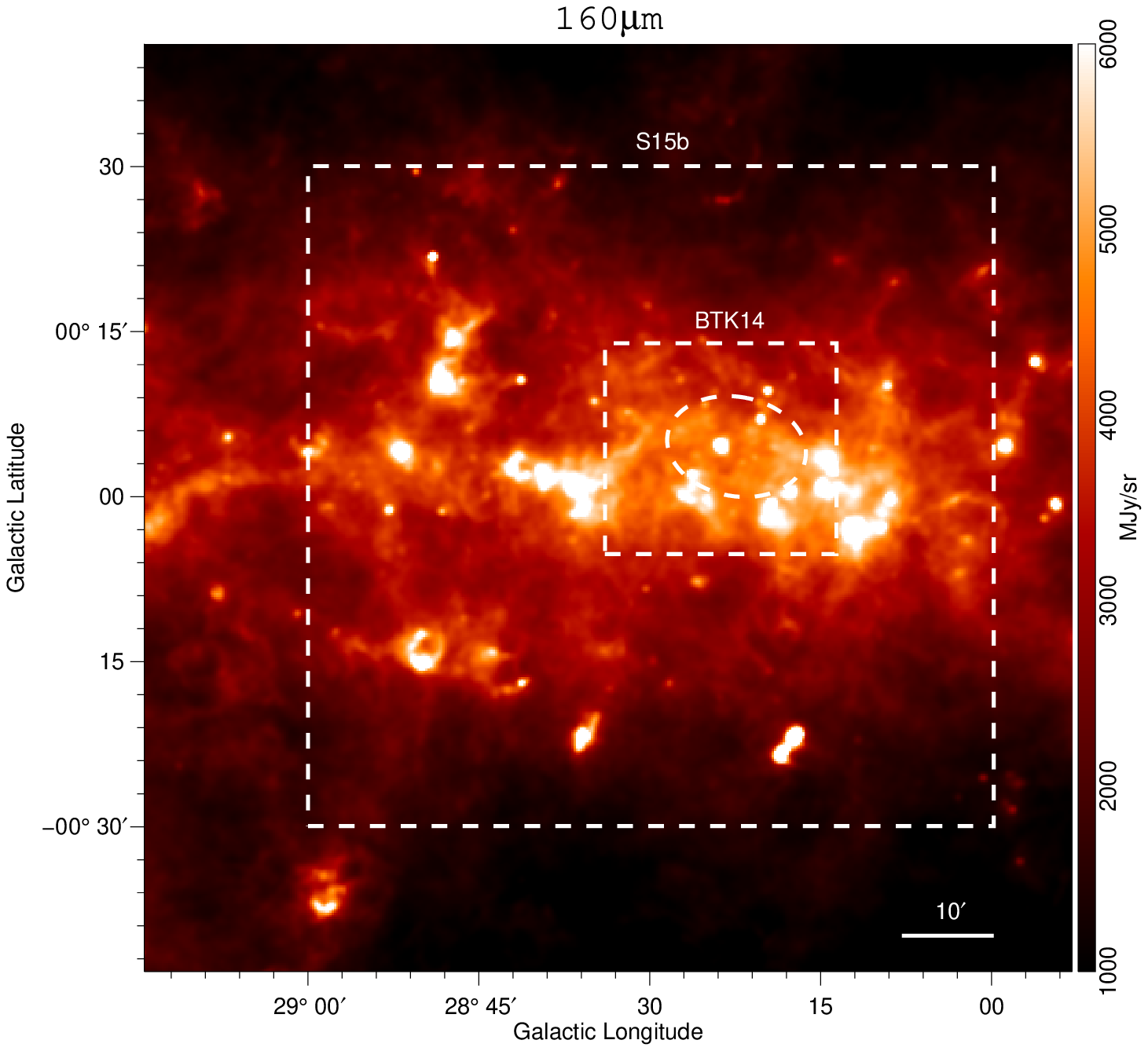}\\
\end{tabular}
\begin{tabular}[b]{c@{\hspace{-0.24in}}c}
\includegraphics[width=5.5in]{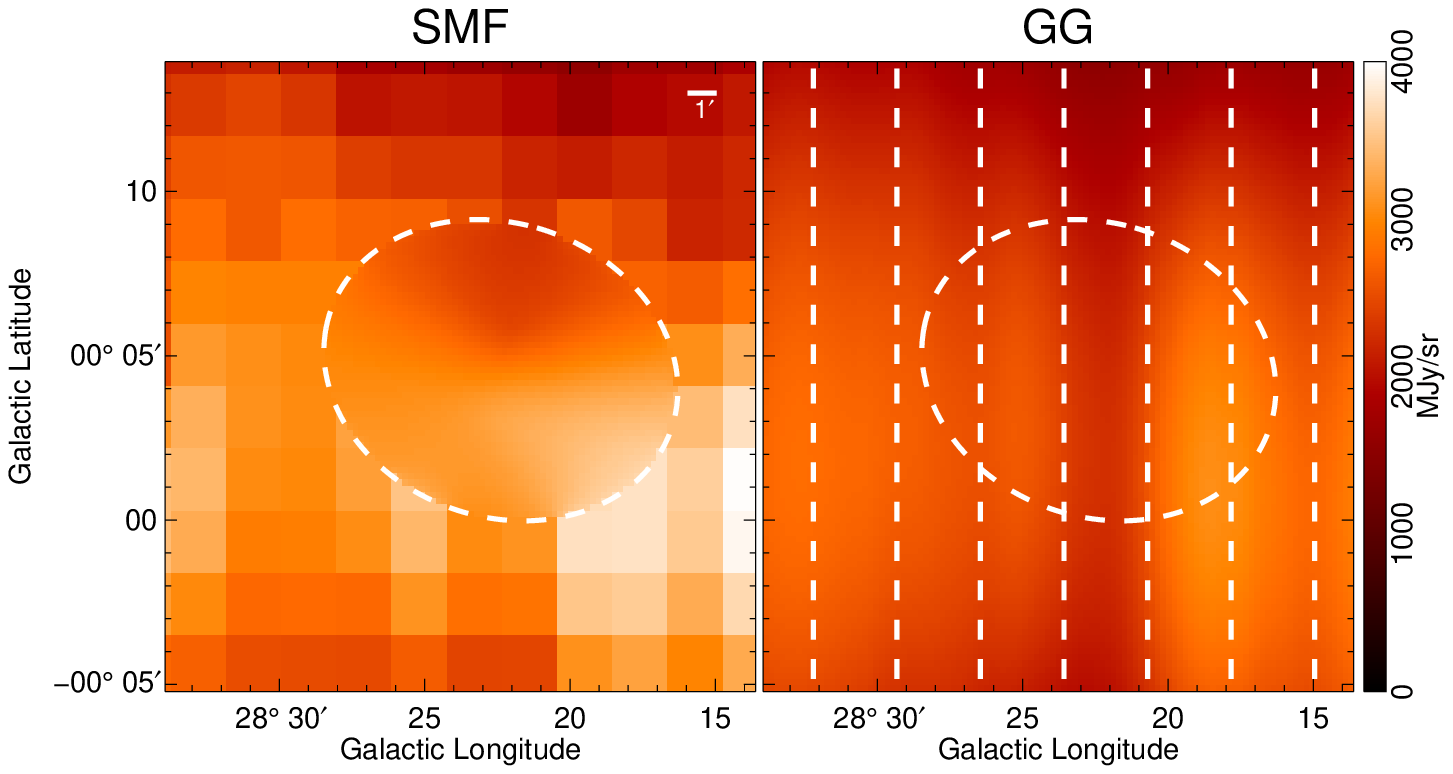} &
\end{tabular}
\caption{\footnotesize 
(a) Top: $160\:{\rm{\mu}{m}}$ image of IRDC G028.37+00.07 (ellipse)
  and surroundings. Smaller rectangle shows $20\arcmin\times19\arcmin$
  area analyzed by BTK14 and in this {\it Letter}. Larger rectangle
  indicates approximate area, $\sim1\degr\times0.7\degr$, studied by
  S15b.
(b) Bottom Left: Galactic background model for $160\:{\rm{\mu}{m}}$
  intensity for the BTK14 region evaluated with Small Median Filter
  (SMF) method. 
(c) Bottom Right: As (b), but with Galactic Gaussian (GG) method. The
  seven vertical lines are locations of profiles shown in Fig.~2.}
\label{fig:fd}
\end{figure*}

Extending these studies to denser, higher-$\Sigma$ clouds, perhaps
more typical of most Galactic star formation, KT13 used combined NIR
(Kainulainen et al. 2011) + MIR (Butler \& Tan 2012 [BT12]) extinction
maps to study $\Sigma$-PDFs of 10 IRDCs. These dense structures,
typically several kpc away, have $\Sigma$-PDFs extending to
$\sim1\:{\rm{g\:cm}^{-2}}$ (e.g., Tan et al. 2014). One uncertainty of
these maps is choice of opacity (i.e., at $8\:{\rm{\mu}{m}}$) per unit
$\Sigma$, with variations $\sim30\%$ expected for different dust
models, e.g., moderately coagulated thin and thick ice mantle models
of Ossenkopf \& Henning (1994 [OH94]). KT13 considered simple,
rectangular regions enclosing contours of
$\Sigma=0.03\:{\rm{g\:cm}^{-2}}$ ($A_V\simeq7\:$mag), i.e.,
``complete'' for $\Sigma$ above this value. However, two caveats limit
this completeness: first, MIR-bright regions are not treated by
extinction mapping, so are excluded from the PDF; second, extinction
mapping saturates at high-$\Sigma$'s, depending on MIR image depth,
typically at $\sim0.3\:$to$\:0.5\:{\rm{g\:cm}^{-2}}$ for Spitzer
GLIMPSE (Churchwell et al. 2009) images (BT12).  At low-$\Sigma$'s,
better probed by NIR extinction mapping, systematic uncertainties are
$\sim0.01\:{\rm{g\:cm}^{-2}}$. The KT13 $\Sigma$-PDFs, extending down
to $A_V=7\:$mag, did not cover PDF peaks well, so provided weak
constraints on true shapes: PDFs could be fit with lognormals or
power laws.

BTK14 presented a higher dynamic range $\Sigma$-map of IRDC
G028.37+00.07, both to higher $\Sigma\simeq0.7\:{\rm{g\:cm}^{-2}}$ and
lower $\Sigma\simeq0.013\:{\rm{g\:cm}^{-2}}$ ($A_V\simeq3\:$mag). From
a $20\arcmin\times19\arcmin$ region around the IRDC (Fig.~1a), BTK14
derived a $\Sigma$-PDF where the peak was observed at
$\Sigma_{\rm{peak}}\simeq0.03\:{\rm{g\:cm}^{-2}}$. Furthermore, the
entire PDF from $0.013\:$to$\:0.4\:{\rm{g\:cm}^{-2}}$ was
well-described by a lognormal with
$\overline{\Sigma}_{\rm{PDF}}=0.039\:{\rm{g\:cm}}^{-2}$
($\overline{A}_{V,{\rm{PDF}}}=9.0\:$mag) and
$\sigma_{\rm{ln}\Sigma^\prime}=1.4$. There appears to be relatively
little mass in a high-$\Sigma$ power law tail, surprising given the
IRDC (KT13) and GMC (Hernandez \& Tan 2015 [HT15]) are both
self-gravitating with virial parameters close to unity and some star
formation has already started.

Another method to measure $\Sigma$ is via sub-mm dust continuum
emission. However, this also depends on dust temperature, $T$, so
multiwavelength studies are needed to probe the spectral energy
distribution (SED) peak. The best data for this comes from Herschel
PACS and SPIRE observations probing
$70\:$to$\:500\:{\rm{\mu}{m}}$. However, derived maps have angular
resolution $\sim20\arcsec$, i.e., $\sim10\times$ worse than the
NIR+MIR extinction maps.

Schneider et al. (2015b [S15b]) utilized Herschel-derived
$\Sigma$-maps to study the same IRDC/GMC examined by BTK14. They
derived $\Sigma$-PDFs for the IRDC ellipse region (Fig.~\ref{fig:fd}a)
and a surrounding ``GMC'' region defined by a $^{13}$CO(1-0) emission
contour equivalent to $A_V\sim2\:$mag, extending approximately over
the larger rectangle shown in Fig.~\ref{fig:fd}. Note, this
is significantly larger than the BTK14 region. S15b found their IRDC
$\Sigma$-PDF was well-fit by a single power-law for
$A_V\gtrsim30\:$mag. The GMC region could also be fit with a power
law, especially for $A_V\gtrsim40\:$mag. Below this S15b claimed to
detect a peak in the $\Sigma$-PDF at $A_V\sim20\:$mag. S15b proposed
that presence of power law tails indicated the cloud was undergoing
multi-scale, including global, quasi-free-fall collapse.

Here we revisit the $\Sigma$-PDF toward IRDC/GMC G028.37+00.07,
especially comparing PDFs derived from dust extinction and emission
methods. We examine reasons for the different results of BTK14 and
S15b and derive the most accurate $\Sigma$-PDF of this massive
protocluster.

\section{Methods}\label{S:method}

\subsection{NIR+MIR Extinction Derived $\Sigma$ Map}

We utilize the NIR+MIR extinction map from BTK14 with one
modification. The NIR extinction map, based on statistical estimates
of stellar extinctions in $\sim30^\prime$ regions, requires choosing
an ``off-position'' of negligible local IRDC/GMC extinction.
BTK14 utilized an off-position at $l=28.4^\circ,\:b=-0.04^\circ$. However, S15b
noted this location may be too close to the GMC, which is confirmed in
the $^{13}$CO(1-0) map of HT15. We therefore choose a new off-position
at $l=28.3^\circ,\:b=+0.3^\circ$ with $A_V$ that is $3\:$mag smaller than the 
previous position. The net effect is to add an offset of $A_V=3\:$mag 
to the BTK14 map. We will see this has only a minor effect on the
$\Sigma$-PDF: in particular, the mean extinction remains close to
$\overline{A}_{V,{\rm{PDF}}}\sim9\:$mag.

Conversion of the extinction map into a $\Sigma$-map requires an
assumption about opacity per unit mass at a given wavelength. Here for
the Spitzer IRAC $\sim8\:{\rm{\mu}{m}}$ map we adopt
$7.5\:{\rm{cm^2\:g}}^{-1}$ (BT12). The NIR+MIR combination assumes
$\tau_{\rm{8{\mu}{m}}}=0.29\tau_K$ (KT13). For conversion to $A_V$ we
follow KT13, adopting
$\Sigma/(1\:{\rm{g\:cm}^{-2}})\equiv{A}_V/(224.8\:{\rm{mag}})$.
Overall, from consideration of different dust models (BT12), we
estimate $\sim30\%$ systematic uncertainties due to these opacity
choices.

\subsection{Sub-mm Emission Derived $\Sigma$ Map}\label{S:mbg}

\begin{figure}
\begin{center}$
\begin{array}{c}
\hspace{-0.1in} \includegraphics[width=2.5in]{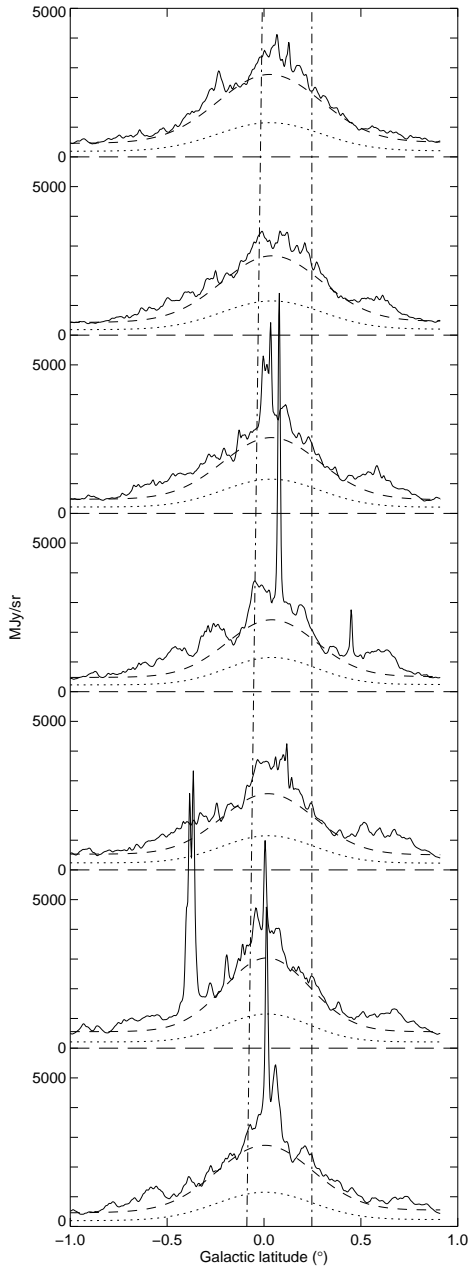}\\
\end{array}$
\end{center}
\caption{\footnotesize 
Seven example profiles of the Galactic Gaussian (GG) background model
(dashed lines), evaluated along $l$ values shown in
Fig.~\ref{fig:fd}c. Solid lines show total observed intensity (after
FG subtraction), dashed lines show fitted Gaussian function to this
profile, and dotted line is GG FG model. Vertical dot-dashed lines
show Galactic latitude range of BTK14 region.}
\label{fig:bg}
\end{figure}

We use PACS and SPIRE images from
the Herschel Infrared GALactic plane survey (Hi-GAL; Molinari et
al. 2010). Derivation of $\Sigma$ and $T$ is performed via
pixel-by-pixel graybody fits to the
$160,\:250,\:350\:$and$\:500\:{\rm{\mu}{m}}$ data (e.g., Battersby et
al. 2011; S15b), first re-gridded at the $500\:{\rm{\mu}{m}}$ image
resolution of 36$\arcsec$.
Specifically, $\Sigma$ and $T$ are derived via:
\begin{equation}
I_{\nu}\simeq{B}_{\nu}(1-e^{-\tau_{\nu}})=B_{\nu}(1-e^{-\Sigma\kappa_{\nu}})
\label{eq:gb}
\end{equation} 
where $I_{\nu}$ is observed intensity of the corresponding band,
$B_{\nu}$ is filter-weighted value of the Planck function,
$\tau_{\nu}$ is optical depth and $\kappa_{\nu}$ is filter-weighted
opacity.

However, to derive local cloud properties, the images need correction
for foreground (FG) and background (BG) diffuse ISM emission along the
line of sight. Assuming emission from the cloud in these sub-mm bands
is optically thin, one method is to estimate FG+BG emission as one
combined, constant intensity. S15b adopted this method,
selecting a region outside the GMC of interest for FG+BG column
density corresponding to $A_V\simeq2\:$mag,
which was then subtracted as a constant offset.

We first assess the FG based on the ISM model derived from observed
$24\:{\rm{\mu}{m}}$ intensities towards ``saturated'' dark regions of
the IRDC by LT14 and Lim, Carey \& Tan (2015 [LCT15]). These saturated
regions, also seen at shorter wavelengths and possibly at
$70\:{\rm{\mu}{m}}$, are where observed intensities are similar to
within instrumental noise in spatially independent locations. This is
assumed to be caused by the IRDC blocking essentially all BG light, so
the observed intensity is that of a spatially smooth FG. Several
$24\:{\rm{\mu}{m}}$ FG measurements across the IRDC are made and then
averaged to estimate a mean. Then the Draine \& Li (2007) diffuse ISM
SED model is normalized to this value to predict
$160,\:250,\:350,\:500\:{\rm{\mu}{m}}$ filter-weighted FG
values. These are subtracted from the sub-mm images, i.e., the FG is
first assumed to be constant across the IRDC/GMC.

Once we have FG-subtracted images, we next assess the BG in two ways.
First, we assess BG intensity at each wavelength adapting the ``Small
Median Filter'' (SMF) method (Butler \& Tan 2009 [BT09]), which is
applied outside the IRDC ellipse with square filter of 35$\arcmin$
size.
BG emission behind the IRDC ellipse is estimated via interpolation
from the surrounding regions (Fig.~1b), following the BT09 weighting
scheme.

As a second, ``Galactic Gaussian'' (GG), method we follow Battersby et
al. (2011) and assume that the Galactic BG follows Gaussian profiles
in latitude. We fit Gaussians to the minimum intensities exhibited
along strips with longitude width $\sim2\arcmin$ (see Fig. 1c and
Fig.~2 for examples at $160\:{\rm{\mu}{m}}$). The method involves
clipping higher intensity values arising from discrete clouds, and
iteratively converges on a final result. Note, here we also assume the
FG intensity is a Gaussian of the same width, and subtract that off
the images before deriving the final BG model. 

The above methods result in FG+BG-subtracted images at
$160,\:250,\:350\:\&\:500\:{\rm{\mu}{m}}$. Then at each pixel we fit
the graybody function (eq.~\ref{eq:gb}), including its filter response
weighting, to derive $\Sigma$ and $T$. This fitting requires an
assumed form of $\kappa_\nu$. This has been studied via MIR to FIR
extinction by LT14 and LCT15, who find evidence of generally flatter
extinction laws over the $\sim8$ to $70\:{\rm{\mu}{m}}$ range as
$\Sigma$ increases, consistent with OH94 and Ormel et al. (2011) dust
models that include grain growth via ice mantle formation and
coagulation. For consistency with these extinction results, we adopt
the OH94 thin ice mantle model with $10^5\:$yr of coagulation at
density of $n_{\rm{H}}=10^6\:{\rm{cm}}^{-3}$ as our fiducial. At
sub-mm wavelengths,
this model exhibits
$\kappa_{\nu}=0.1\:(\nu/1000\:{\rm{GHz}})^{\beta}\:{\rm{cm}}^2\:{\rm{g}}^{-1}$
with $\beta\simeq1.8$. We will also explore the effects of varying
$\beta$ from 1.5 to 2.

Finally, we generate higher resolution (HiRes) $\Sigma$ maps by
re-gridding to the $250\:{\rm{\mu}{m}}$ image pixels (18$\arcsec$
resolution; 6$\arcsec$ pixels) and then repeating the above analysis,
but now fixing temperatures from the lower angular resolution
maps. These HiRes $\Sigma$ maps are better able to probe smaller,
higher $\Sigma$ structures.

\section{Results}

\subsection{Mass surface density and temperature maps}

\begin{figure*}
\begin{center}$
\begin{array}{cc}
\hspace{-0.1in} \includegraphics[width=6.5in]{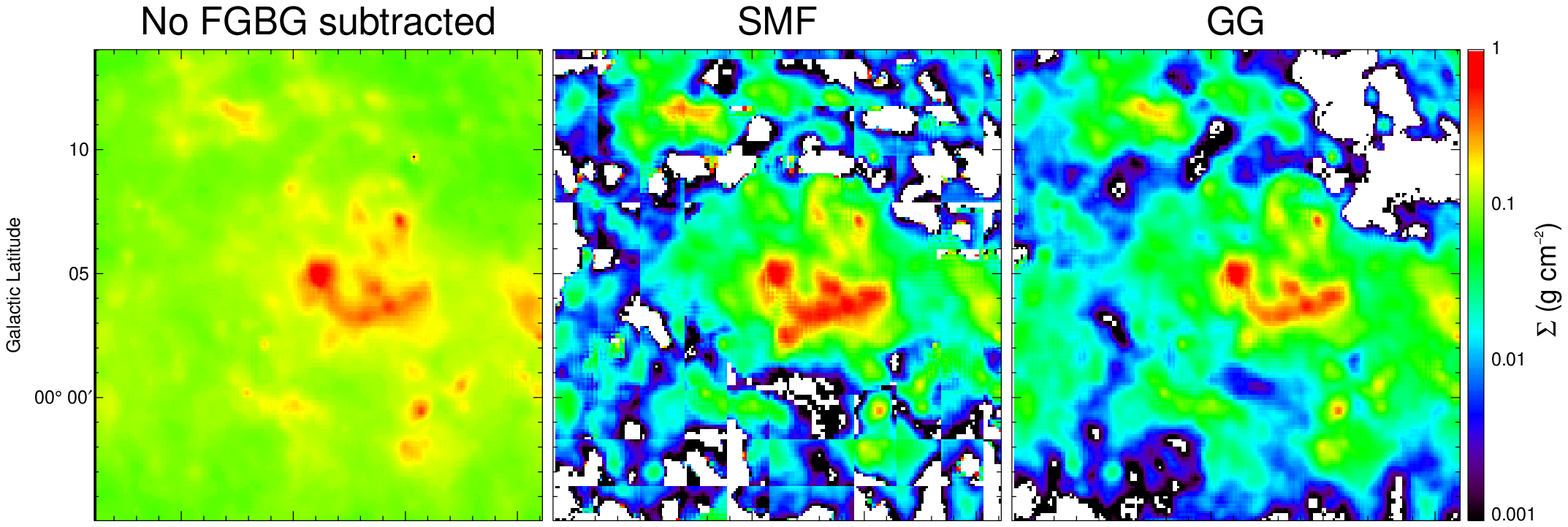}  \\
\hspace{-0.1in} \includegraphics[width=6.5in]{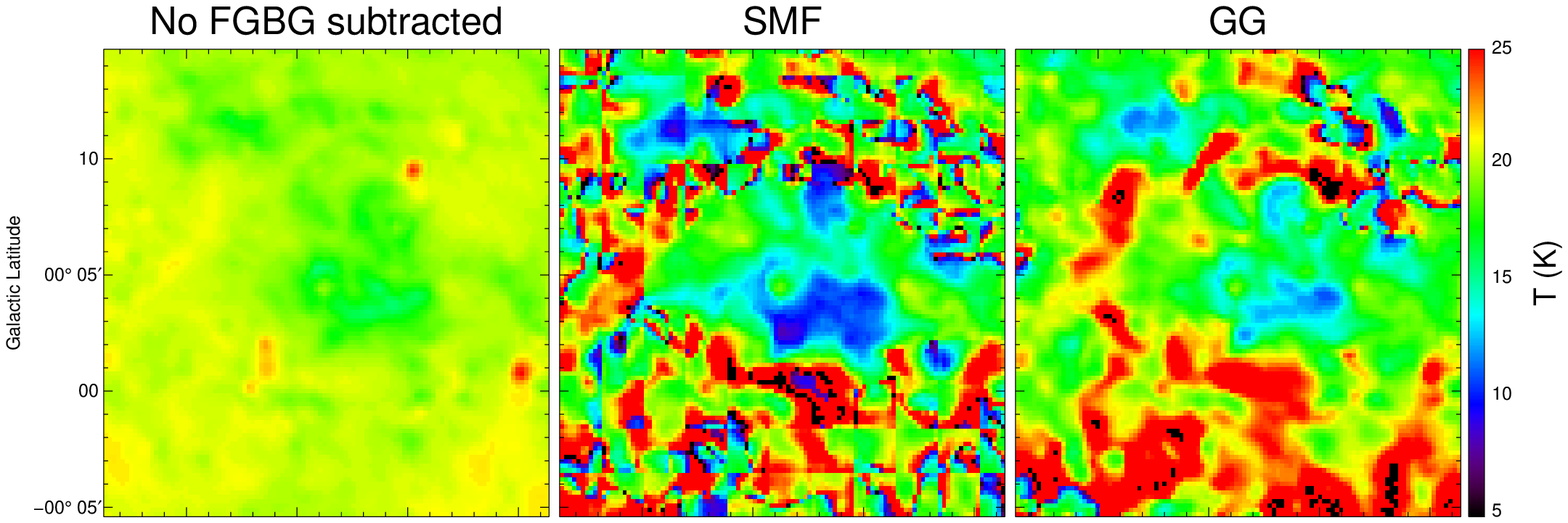}  \\
\hspace{-0.1in} \includegraphics[width=6.5in]{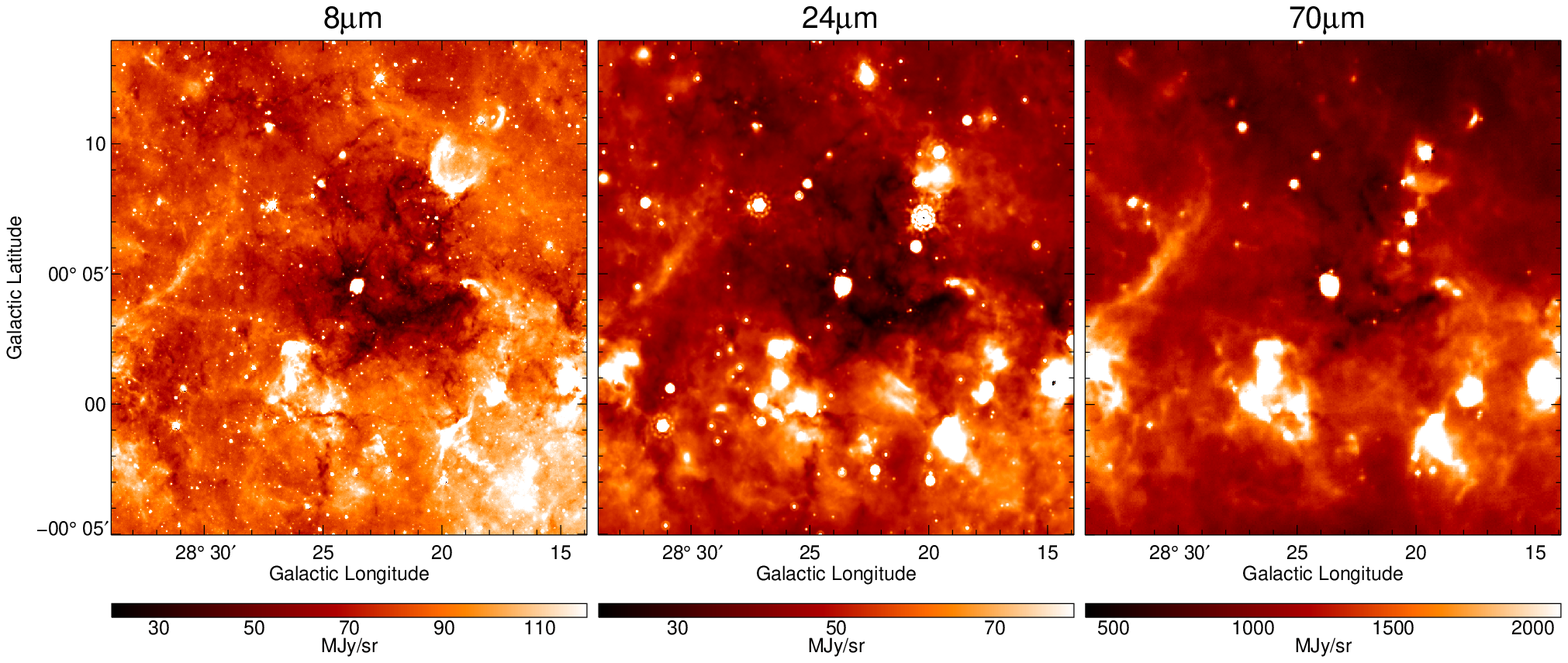}  \\
\end{array}$
\end{center}
\caption{\footnotesize 
Sub-mm emission derived HiRes $\Sigma$ (top row) and $T$ (middle row) maps
based on no FG+BG subtraction (left), SMF (middle) and GG (right)
methods. Bottom row shows images from Spitzer-IRAC
$8\:{\rm{\mu}{m}}$, -MIPS $24\:{\rm{\mu}{m}}$ and Herschel-PACS
$70\:{\rm{\mu}{m}}$.}
\label{fig:ST}
\end{figure*}

Figure~\ref{fig:ST} shows the sub-mm emission derived $\Sigma$ (HiRes)
and $T$ maps of the IRDC, starting from maps derived with no FG and BG
subtraction, and then showing the effects of the two background
estimation methods (SMF and GG). The overall result of FG+BG
subtraction on the $\Sigma$ map is to effectively remove
$\sim0.1\:{\rm g\:cm^{-2}}$ across the cloud. As we will see this has
a major effect on the shape of the $\Sigma$-PDF. Note for study of the
BTK14 region, we consider the GG method to be superior to SMF as it
allows for the large-scale structure of the Galactic plane. Also, the
SMF method tends to underestimate mass surface densities in the
structures just outside the IRDC ellipse.

The temperature maps are also strongly affected by FG+BG subtraction,
which leads to a lowering of the temperatures measured in the IRDC, as
well as revealing warmer localized patches in the surroundings. Some
of these appear to correspond to
$8,\:24\:\&\:70\:{\rm{\mu}{m}}$-brighter regions (Fig.~3 lower
panels).

The top row of Figure~4 compares the FG+BG-subtracted $\Sigma$ maps
derived from NIR+MIR extinction and from sub-mm emission. White
patches in the extinction map are locations of bright MIR emission
that prevent an absorption measurement against the Galactic
background. White patches in the sub-mm emission derived map are
locations where the background subtraction has caused the estimated
flux from the cloud to become negative in at least one wavelength.
The second row of Fig.~4 presents the same information, but now with a
simplified color scheme for the $\Sigma$ scalebar, which can be
compared to regions of the $\Sigma$-PDF, discussed below.

Broadly similar morphologies are seen in these maps, but with the
sub-mm emission derived maps tending to find moderately higher values
in the IRDC, although the differences between the SMF and GG
background subtraction methods are comparable to the differences
between the Sub-mm Em. (GG) model and the extinction map. Note also
the superior resolution of the extinction map.

Fig.~4's third row shows the effect of applying different values of
$\beta$ for deriving $\Sigma$ (GG case). Note $\beta=1.8$ is closest
to the behavior of the OH94 thin ice mantle model. Slightly lower
values of $\beta\simeq1.5$ (which lead to warmer derived temperatures
and thus lower values of $\Sigma$; e.g., Guzm\'an et al. 2015) are one
way of reconciling differences between the extinction and emission
derived maps.

Finally, Fig.~4 also shows a pixel-by-pixel comparison of $\Sigma$'s
derived by NIR+MIR extinction and the fiducial sub-mm emission (GG)
method. The fraction of the pixels with both $\Sigma$'s
$>0.013\:{\rm{g\:cm}}^{-2}$ with values within 30\% of each other is 0.608.

\begin{figure*}
\begin{center}$
\begin{array}{cc}
\hspace{-0.1in} \includegraphics[width=6.5in]{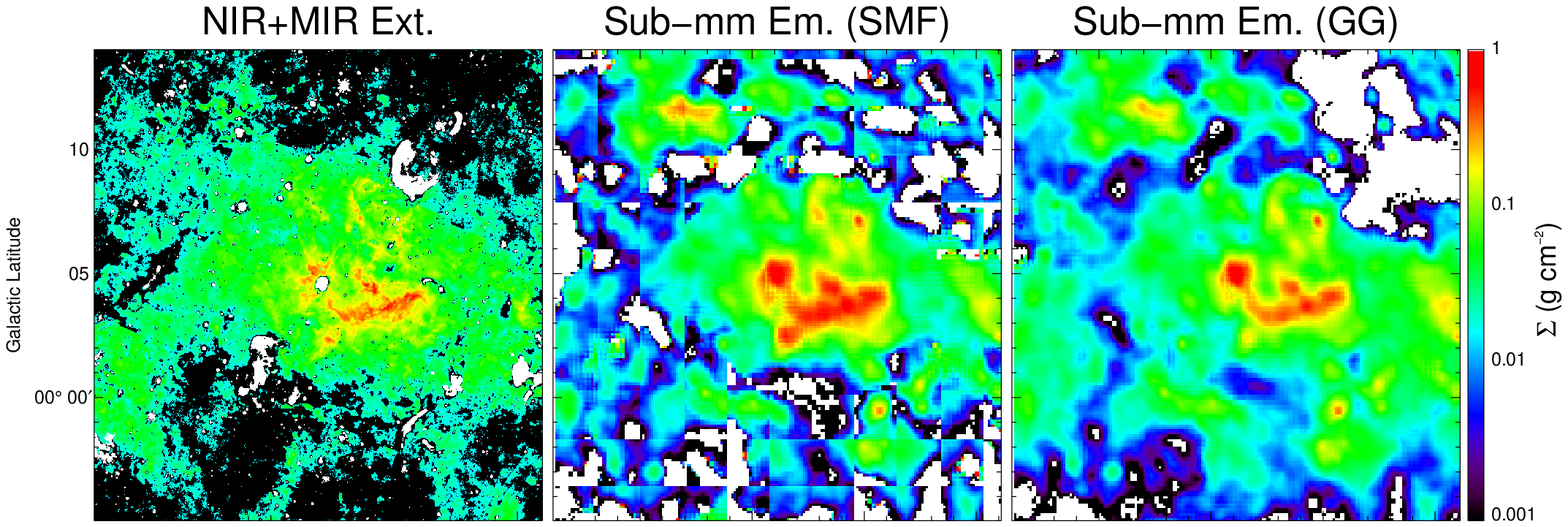}  \\
\hspace{-0.1in} \includegraphics[width=6.5in]{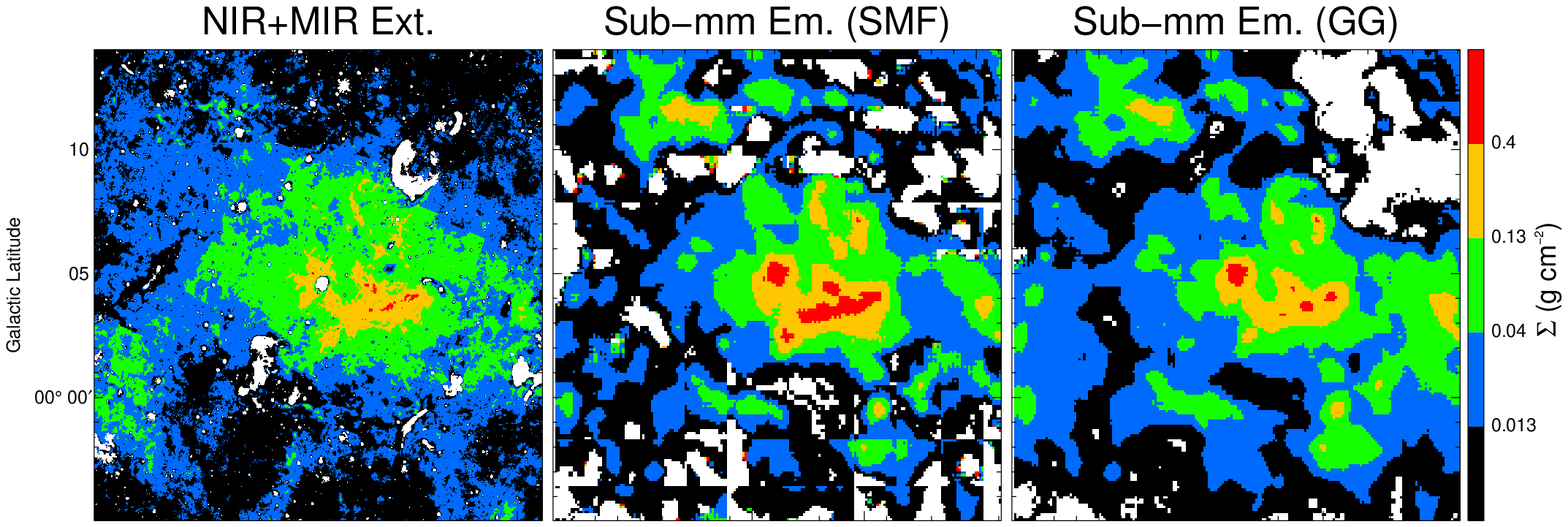}  \\
\hspace{-0.1in} \includegraphics[width=6.5in]{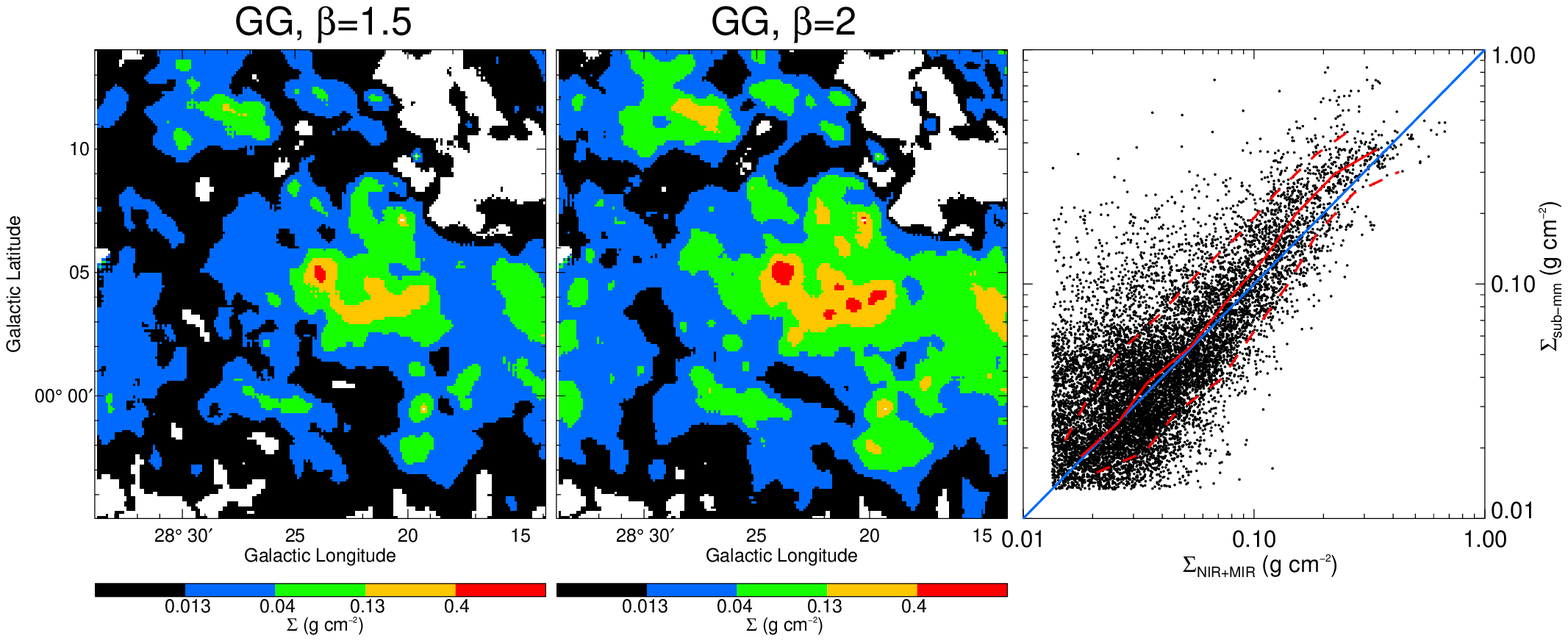}  \\
\end{array}$
\end{center}
\caption{\footnotesize 
{\it{Top row, left-to-right:}} NIR+MIR extinction derived $\Sigma$ map,
sub-mm emission derived $\Sigma$ map using SMF BG subtraction,
sub-mm emission derived $\Sigma$ map using GG BG subtraction.
{\it{Middle row, left-to-right:}} As top row, but with simplified
color scheme for $\Sigma$ scalebar to highlight values around
peak of the $\Sigma$-PDFs.
{\it{Bottom row, left-to-right:}} Sub-mm emission derived $\Sigma$ map
(GG BG subtraction), with $\beta=1.5,\:2.0$, and pixel by pixel
$\Sigma$ comparison between NIR+MIR extinction and fiducial sub-mm
emission (GG) methods (red lines show median and $1\sigma$ range in
data binned perpendicular to the one-to-one line).}
\label{fig:Sigma}
\end{figure*}

\subsection{$\Sigma$ Probability Distribution Functions}
\label{S:ppdf} 

\begin{figure*}
\begin{center}$
\begin{array}{c}
\hspace{-0.1in} \includegraphics[width=7.2in]{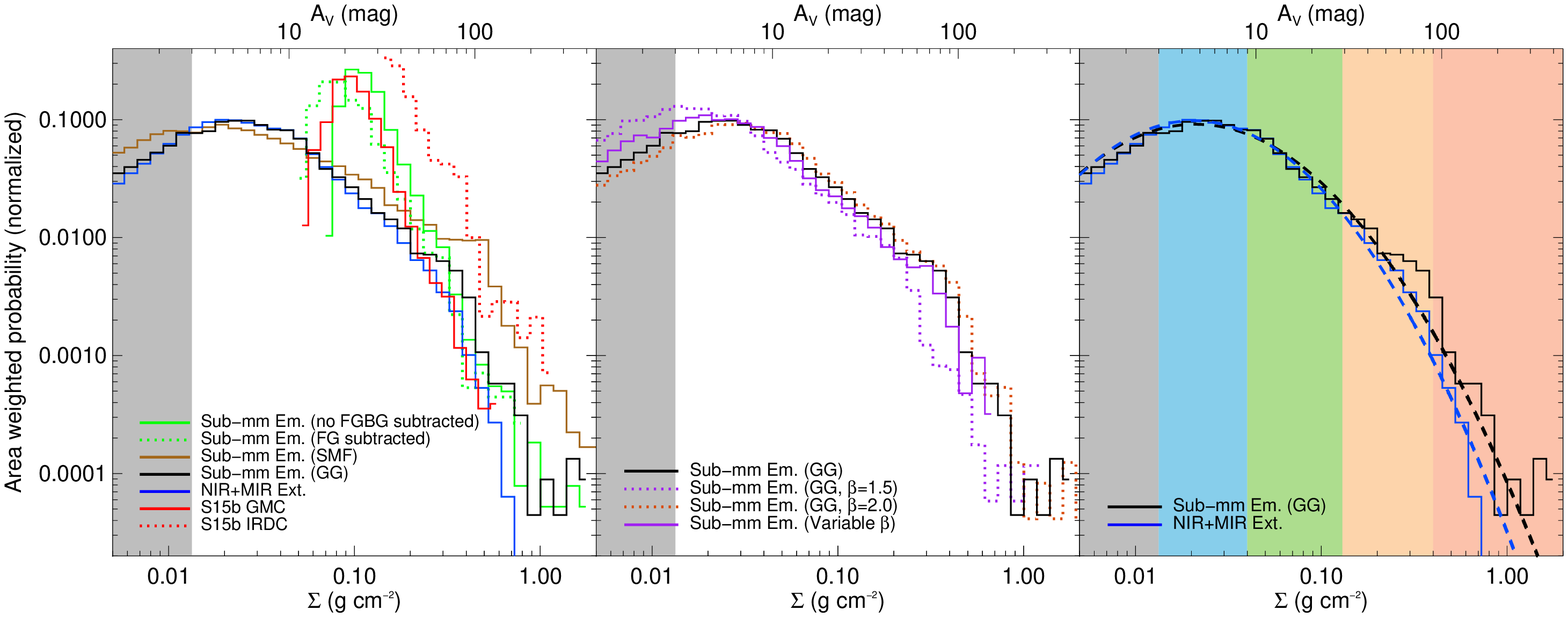} \\
\hspace{-0.1in} \includegraphics[width=7.2in]{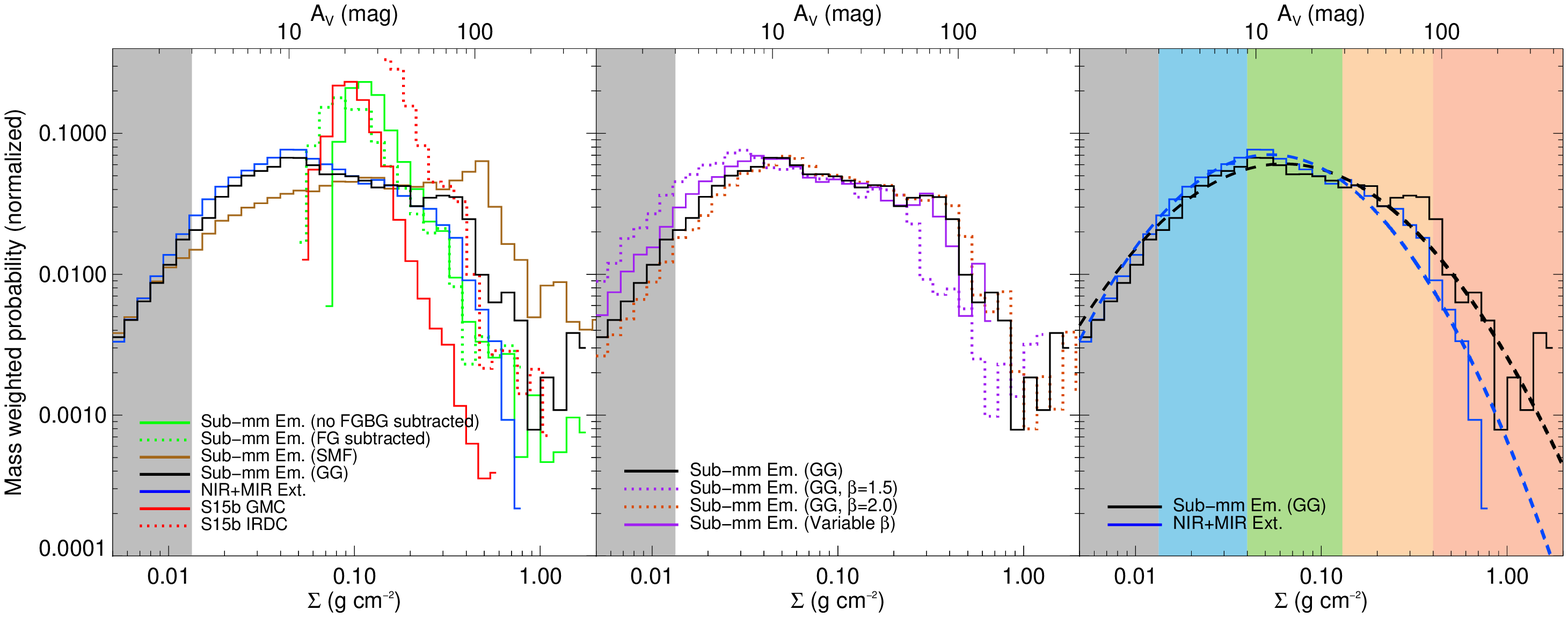} \\
\end{array}$
\end{center}
\caption{\footnotesize 
Top Row: Area-weighted $\Sigma$-PDFs of IRDC/GMC G028.37+00.07. (a)
Left: Effects of foreground/background subtraction. Green solid line
shows Sub-mm Em. case with no FG/BG subtracted. Green dotted line
shows only FG subtracted. Brown/black solid shows SMF/GG BG
subtraction, respectively. Blue solid line shows NIR+MIR extinction
result, including FG+BG subtraction.
(b) Middle: Effects of $\beta$ variation. Black solid line shows the
fiducial Sub-mm Em. GG case, which has effective $\beta\simeq1.8$. The
$\beta=1.5,\:2$ cases are shown with dotted purple and orange lines,
respectively. Variable $\beta$ model is shown with a solid purple
line.
(c) Right: Lognormal fits to fiducial $\Sigma$-PDFs. Shaded regions
correspond to colors shown in Fig.~4 middle and bottom rows. Note grey
region shows values below minimum closed contour.
Bottom row, (d), (e), (f): As top row, but for mass-weighted
$\Sigma$-PDFs.}
\label{fig:pdf}
\end{figure*}

Figure~5a shows the area-weighted $\Sigma$-PDF of IRDC/GMC
G028.37+00.07. Note PDF normalization is with respect to the total
``BTK14'' area (see Fig.$\:1$). The result from the sub-mm dust
emission map with no FG and BG subtracted (green solid line) is
similar to S15b's results, being intermediate between their GMC and
IRDC PDFs (normalized to their particular areas), as expected given
the region geometries (Fig.$\:1$). FG-only subtraction (green dotted)
has a modest effect: the PDF peak moves to lower $\Sigma$ by
$\simeq0.02\:{\rm{g\:cm}}^{-2}$.

Background subtraction (SMF: brown solid; GG: black solid) leads to
larger shifts of the PDF peak, reducing its value by factors of
several.  Overall the shapes of the two sub-mm emission derived
$\Sigma$-PDFs are quite similar. The NIR+MIR derived PDF (blue solid)
is also similar, especially to the PDF using GG background estimation.

Figure~5b illustrates effects of varying $\beta$ from 1.5 to 2 on the
Sub-mm Em. (GG) PDF. Smaller values of $\beta$ reduce the amount of
inferred high-$\Sigma$ material (thus boosting the low-$\Sigma$
distribution). Dust properties may vary systematically with $\Sigma$
due to grain growth (LCT15). Figure~5b also shows a simple
``Variable-$\beta$'' model where
$\beta=1.8-0.3(\Sigma/[0.5\:{\rm{g\:cm}^{-2}}])$, i.e., lower $\beta$
in high-$\Sigma$ regions broadly consistent with models of grain
growth (e.g., Ormel et al. 2011), applied iteratively until
convergence is achieved. The resulting $\Sigma$-PDF is quite similar
to the fiducial Sub-mm Em. (GG) case.

Figure~5c shows lognormal fits to the Sub-mm Em. (GG) and NIR+MIR
Ext. derived PDFs, with
$\Sigma_{\rm{peak}}\simeq0.023,\:0.019\:{\rm{g\:cm}^{-2}}$
($A_V\simeq5.1,\:4.3\:$mag), respectively. Note, fitting is done only
for $\Sigma\geq0.013\:{\rm{g\:cm}^{-2}}$ ($A_V\simeq3\:$mag), i.e.,
above the minimum closed contour in the NIR+MIR extinction map.
With $\Sigma_{\rm{peak}}$ greater than $\Sigma$ of the minimum closed contour,
the lognormal fitting is well-constrained by the data. Values of
$\overline{\Sigma}_{\rm PDF}$ are
$0.044,\:0.038\:{\rm{g\:cm}^{-2}}$, while
$\sigma_{\rm{ln}\Sigma^\prime}=1.35,\:1.15$, respectively, compared to
1.4 reported by BTK14.

The $\Sigma$-PDFs are well-fit by single lognormals. The fraction of
mass above $\Sigma=0.013\:{\rm{g\:cm}^{-2}}$ that is in excess of the
lognormal fits in their high-$\Sigma$ ($>0.1\:{\rm{g\:cm}^{-2}}$)
tails is $\epsilon_{\rm{pl}}\sim0.08,\:0.03$ for Sub-mm Em. (GG) and
NIR+MIR Ext. PDFs, respectively. Such small fractions may be
consistent with similarly low values of $\epsilon_{\rm{ff}}$. Krumholz
\& Tan (2007) estimated $\epsilon_{\rm{ff}}\sim0.02$, including
results from observed IRDCs, while Da Rio et al. (2014) estimated
$\epsilon_{\rm{ff}}\simeq0.04$ in the Orion Nebula Cluster. However,
detailed study of numerical simulations to link the mass fraction in
these ``tails,'' i.e., $\epsilon_{\rm{pl}}$, with $\epsilon_{\rm{ff}}$
is still needed for trans-Alfv\'enic, turbulent, global clouds.

Note, these high-$\Sigma$ excess fractions in the PDFs are in fact not
particularly well-described with power law tails to the
lognormals. Given the small size of the excess fractions, the limited
dynamic range of $\Sigma$ where they appear, and the potential
systematic errors that enter at high $\Sigma$'s, we do not fit power
law functions, but rather focus on $\epsilon_{\rm{pl}}$ as our metric
for deviation of the PDF from a lognormal shape.

Note also, while we consider the GG method preferable to SMF,
log-normal fitting results are not too sensitive to this choice: with
SMF $\Sigma_{\rm{peak}}$ decreases by 25\%,
$\sigma_{\rm{ln}\Sigma^\prime}$ increases by 25\%, and
$\epsilon_{\rm{pl}}$ decreases by 20\%.

Figures 5d-f mirror Figs.~5a-c, but now for mass-weighted PDFs. Values
of $\Sigma_{\rm{peak}}$ of the lognormal fits are
$0.060,\:0.051\:{\rm{g\:cm}^{-2}}$ for the Sub-mm Em. (GG) and NIR+MIR
Ext. PDFs, respectively. Values of $\overline{\Sigma}_{\rm{PDF}}$ are
$0.067,\:0.054\:{\rm{g\:cm}^{-2}}$, 
while $\sigma_{\rm{ln}\Sigma^\prime}\simeq1.10,\:0.96$, respectively.
These mass-weighted $\Sigma$-PDFs are also well-fit by single
lognormals, with the PDF peak being significantly above the minimum
closed contour level.

\section{Discussion}

We have measured the $\Sigma$-PDF from NIR+MIR extinction and sub-mm
dust emission in a contiguous $30\:$pc-scale region centered on a
dense, massive IRDC that extends to its surrounding GMC. This material
is likely to eventually form a massive star cluster. The two methods
give similar results, especially detecting the area-weighted
$\Sigma$-PDF peak at $\simeq0.021\:{\rm{g\:cm}^{-2}}$ and the
mass-weighted $\Sigma$-PDF peak at $\simeq0.055\:{\rm{g\:cm}^{-2}}$,
both significantly higher than the minimum closed contour at
$\simeq0.013\:{\rm{g\:cm}^{-2}}$. Comparison of extinction and
emission-based methods is important to assess systematic
uncertainties. The consistency of these results increases our
confidence in the reliability of the $\Sigma$ maps and their PDFs.

Some differences may result from dust opacity uncertainties, including
potential systematic variations with density due to grain growth,
which have greater influence on the sub-mm emission method. Angular
resolution also leads to differences: the NIR+MIR extinction map has
$\sim2$\arcsec\ resolution, while the sub-mm emission derived map has
18\arcsec\ resolution, which will tend to smooth out high-$\Sigma$
peaks, thus limiting its ability to see highest-$\Sigma$ regions.
NIR+MIR extinction mapping faces problems of saturation at high
$\Sigma$, but this should only become important at
$\Sigma\gtrsim0.65\:{\rm{g\:cm}}^{-2}$ (BTK14): i.e., most of the
range of $\Sigma$ of Fig.~5 is unaffected. NIR+MIR extinction mapping
also fails in MIR-bright regions. Fig.~4 middle row and Fig.~5c show
that a significant reason for difference in the amount of
$\Sigma>0.4\:{\rm{g\:cm}}^{-2}$ material in the NIR+MIR extinction map
compared to the Sub-mm Em. (GG) map results from the material around
the central MIR-bright source in the IRDC. However, this is not enough
to explain the claimed power law tail of S15b's analysis.
Rather, most of the difference of our results from  
S15b's are caused by our higher estimate of the diffuse Galactic plane
background subtraction level.

The $\Sigma$-PDFs are well-fit by single lognormals, even though this
IRDC and GMC region is gravitationally bound with virial parameters of
about unity (KT13; HT15). The $\Sigma$-PDF peak is greater than the
minimum closed contour: i.e., peak position is not too sensitive to
choice of map boundary. If we analyze a smaller
$15\arcmin\times15\arcmin$ region centered on the IRDC, then
$\Sigma_{\rm{peak}}\simeq0.023\:{\rm{g\:cm}^{-2}}$,
$\sigma_{\rm{ln}\Sigma^\prime}=1.22$ and
$\epsilon_{\rm{pl}}\simeq0.07$ (averaging results from Sub-mm Em. (GG)
and NIR+MIR extinction methods), similar to the results for the
$20\arcmin\times19\arcmin$ region.  The position of this peak likely
has physical significance, e.g., depending on properties of
protocluster turbulence, magnetic fields and/or feedback, e.g.,
protostellar outflows, and thus constrains theoretical models of star
cluster formation.

Another important result is the mass fraction in the high-$\Sigma$
power law tail (or lognormal excess), $\epsilon_{\rm{pl}}$. There is
tentative evidence for a small tail being present at
$\Sigma\gtrsim0.2\:{\rm{g\:cm}}^{-2}$ and containing
$\sim3\:$to$\:8\%$ of the total mass, but subject to the systematic
uncertainties discussed above. Still, we consider this to be the most
accurate measurement of this high-$\Sigma$ lognormal excess mass
fraction since we have measured $\Sigma$ with two independent methods,
which both detect the lognormal peak. This mass fraction also
constrains theoretical models, especially protocluster star formation
rate and thus duration of star cluster formation. Relatively small
$\epsilon_{\rm{pl}}$ may imply small $\epsilon_{\rm{ff}}$ (Krumholz \&
McKee 2005; Kritsuk et al. 2011; Federrath \& Klessen 2012), and thus
an extended duration for star cluster formation (Tan et al. 2006; Da
Rio et al. 2014). Better quantification of the relation between
$\epsilon_{\rm{pl}}$ and $\epsilon_{\rm{ff}}$ should be an additional
goal of star cluster formation simulations.

\acknowledgements JCT acknowledges support from NASA ADAP grant NNX15AF21G.


\begin{thebibliography}{}
\bibitem[Battersby et al.(2011)]{battersby11} Battersby, C., Bally, J., Ginsburg, A., et al. 2011, \aap, 535, 128 
\bibitem[Butler \& Tan(2009)]{butler2009} Butler, M. J. \& Tan, J. C., 2009, \apj, 696, 484
\bibitem[Butler \& Tan(2012)]{butler2012} Butler, M. J. \& Tan, J. C., 2012, \apj, 754, 5
\bibitem[Butler, Tan \& Kainulainen(2014)]{butler2014} Butler, M. J., Tan, J. C. \& Kainulainen, J., 2014, \apj, 782L, 30
\bibitem[Cho \& kim2011]{cho11} Cho, W. \& Kim, J., 2011, \mnras, 410, 8
\bibitem[Churchwell et al.(2009)]{churchwell2009} Churchwell, E., Babler, B., Meade, M., et al. 2009, PASP, 121, 213
\bibitem[Collins et al.(2011)]{collins11} Collins, D. C., Padoan, P., Norman, M. L., \& Xu, H. 2011, \apj, 731, 59 
\bibitem[Da rio et al.(2014)]{dario14} 	Da Rio, N., Tan, J. C., Jaehnig, K. 2014, \apj, 795, 55
\bibitem[Draine \& Li(2007)]{draine2007} Draine, B. T. \& Li, A. 2007, \apj, 675, 810
\bibitem[Federrath(2013)]{federrath13} Federrath, C. 2013, \mnras, 436, 1245
\bibitem[Federrath \& Klessen(2012)]{federrath_klessen12} Federrath, C. \& Klessen, R. S. 2012, \apj, 761, 156
\bibitem[Federrath \& Klessen(2013)]{federrath_klessen13} Federrath, C. \& Klessen, R. S. 2013, \apj, 763, 51
\bibitem[Guzm\'{a}n et al. 2015]{guzman15} Guzm\'{a}n, A., Sanhueza, P., Contreras, Y., et al. 2015, \apj, 815, 130 
\bibitem[Hernandez \& Tan(2015)]{hernandez15} Hernandez, A. K. \& Tan, J. C. 2015, \apj, 809, 154
\bibitem[Kainulainen et al.(2009)]{K09} Kainulainen, J., Beuther, H., Henning, T., Plume, R. 2009, \aap, 508, 35
\bibitem[Kainulainen \& Tan(2013)]{KT13} Kainulainen, J. \& Tan, J. C. 2013, \aap, 549, 53
\bibitem[Kritsuk et al.(2011)]{kritsuk11} Kritsuk, A. G., Norman, M. L. \& Wagner, R. 2011, \apj, 727, 20
\bibitem[Krumholz \& Mckee(2005)]{krumholz05} Krumholz, M. R. \& McKee, C. F. 2005, \apj, 630. 250
\bibitem[Krumholz \& Tan(2007)]{krumholz07} Krumholz, M. R. \& Tan, J. C. 2007, \apj, 654, 304
\bibitem[Lim \& Tan(2014)]{lim2014} Lim, W. \& Tan, J. C. 2014, \apj, 780,29
\bibitem[Lim, Carey \& Tan(2015)]{lim2015} Lim, W., Carey, S. J. \& Tan, J. C. 2015, \apj, 814,28
\bibitem[Lombardi et al.(2015)]{lombardi15} Lombardi, M, Alves, J. \& Lada, C. J. 2015, \aap, 576, 1 
\bibitem[Molinari et al.(2010)]{molinari10} Molinari, S., Swinyard, B., Bally, J., et al. 2010, \pasp, 122, 314
\bibitem[Ormel et al.(2011)]{ormel2011} Ormel, C. W., Min, M., Tielens, A. G. G. M, et al. 2011, \aap, 532, 43
\bibitem[Ossenkopf \& Henning(1994)]{ossenkopf1994} Ossenkopf, V. \& Henning, Th. 1994, \aap, 291, 943
\bibitem[Padoan et al.(2014)]{padoan14} Padoan, P., Haugb$\o$lle, T., \& Nordlund, ${\rm \AA}$. \apj, 797, 32
\bibitem[Schneider et al.(2015a)]{S15a} Schneider, N., Ossenkopf, V., Csengeri, T., et al. 2015a, \aap, 575, 79
\bibitem[Schneider et al.(2015b)]{S15b} Schneider, N., Csengeri, T., Klessen, R. S., et al. 2015b, \aap, 578, 29
\bibitem[Tan et al.(2006)]{tan06} Tan, J. C., Krumholz, M. R. \& McKee, C. F. 2006, \apj, 641, 121
\bibitem[Tan et al.(2014)]{tan14} Tan, J. C., Beltr$\acute{a}$n, M. T., Caselli, P., et al. 2014, PPVI, p149, arXiv:1402.0919 

\end{thebibliography}
\end{document}